\documentstyle[12pt]{article} 

 1                                        
\def\vec#1{\mbox{\protect\boldmath $ #1 $}}

\begin{document}                                                              
\begin{flushright}
\begin{minipage}[t]{12em}
UAB-FT-458/98 \\
IFT-P.086/98
\end{minipage}
\end{flushright}                                                      
\vskip 1.5cm
 
\begin{center}                                                                
{ \large\bf BELL--INEQUALITIES FOR ENTANGLED PAIRS OF NEUTRAL KAONS}
\vskip 2cm  

{ Albert Bramon $^{a)}$ \footnote{e-mail address: bramon@ifae.es} and 
Marek Nowakowski $^{a,b)}$ \footnote{e-mail address: marek@ift.unesp.br}} \\
\vskip 0.5cm   
    a) Grup de F{\'\i}sica Te\`orica, \\  
Universitat Aut\`onoma de Barcelona, 08193 Bellaterra, Spain \\
    b) Instituto de F{\'\i}sica Te\'{o}rica, Universidade Estadual Paulista, \\
Rua Pamplona, 145, 01405 -900 S\~ao Paulo, Brazil 
\end{center}

\vskip 6cm                             
\begin{abstract} 

We extend the use of Bell--inequalities to $\Phi \to K^0 \bar{K^0}$ decays 
by exploiting analogies and differences to the well--known and experimentally 
verified singlet--spin case. Contrasting with other analyses, our 
Bell--inequalities are violated by quantum mechanics and can strictly be derived 
from local realistic theories. In principle, quantum mechanics could then be
tested using unstable, oscillating states governed by a CP--violating Hamiltonian.
\end{abstract}                                                                

\newpage

Quantum entanglement, as shown by the separate parts of  
non--factorizable composite systems, 
is an extremely peculiar feature of quantum mechanics to which much attention has been 
devoted. Since the paper by Einstein, Podolsky and Rosen \cite{epr}, quantum
entanglement has been also a continous source of 
speculations on the ``spooky action-at-a-distance'', better characterized as
non-locality in the correlations of an EPR--pair
\cite{peres}. Well--known and useful tools to probe into this non-locality are the  
original Bell--inequalities \cite{bell} and their reformulated versions 
\cite{clh}-\cite{wigner}, as reviewed, for instance, in 
\cite{review1} and \cite{redhead}.

Bell--inequalities have been subjected to experimental tests with the general
outcome that they are violated \cite{bellexp}-\cite{tittel}, i.e., local realistic
theories fail and nature is indeed non--local. However, possible loopholes 
in the tests have been pointed out \cite{santos}. There is therefore 
a continous interest to test Bell--inequalities in different experiments
and, more importantly, in different branches of physics. 
One such possible place is an 
$e^+e^-$ machine copiously producing EPR-entangled $K^0\bar{K^0}$  
pairs through the reaction $e^+e^- \to \Phi \to K^0 \bar{K^0}$. Such a $\Phi$-- or
``entanglement"--factory,  Da$\Phi$ne, will be soon operating in Frascati \cite{handbook}.
Due to the negative charge conjugation of the $\Phi$--meson, the EPR entanglement
of the neutral  kaon pair can be explicitly written as
\begin{eqnarray} \label{1}
|\Phi (0)\rangle &=& {1 \over \sqrt{2}}\left[|K^0\rangle \otimes|\bar{K^0}\rangle
-|\bar{K^0}\rangle \otimes|K^0 \rangle \right].
\end{eqnarray}
Starting with this initial state, the two neutral kaons --denoted by the kets
at the left and at the right hand side of the direct product symbol $\otimes$ in
(\ref{1})-- fly apart thus defining after collimation a left and a right hand kaon beam. 
Their time evolution is given by (see Appendix and \cite{remark1}) 
\begin{eqnarray} \label{1prime}
|\Phi (t) \rangle  &=& {N(t) \over \sqrt{2}} \left[|K_S\rangle
\otimes|K_L\rangle -|K_L\rangle \otimes|K_S \rangle \right]
\end{eqnarray}
with $| N(t) | = (1+|\epsilon |^2) / (|1- \epsilon |^2) 
\times e^{-\frac{1}{2}(\Gamma_S + \Gamma_L) t}$ 
reflecting the extinction of the beams via weak kaon decays but without 
modifying the perfect antisymmetry of the initial state. This is then  
in close analogy to the singlet--spin state usually considered in the
Bohm reformulation of the EPR configuration (EPRB). But there is also a substantial
difference: while most of the experimental tests favouring quantum non--locality have been
successfully performed in the EPRB configuration, early \cite{datta1}-\cite{ghi} 
and more recent \cite{domenico}-\cite{uchiyama} attempts to check similar
Bell--inequalities in $e^+e^- \to \Phi \to K^0 \bar{K^0}$ either fail in showing their
violation by quantum mechanics or seem to be affected by serious difficulties (see 
below and \cite{abn}). Our purpose in this letter is to fill this gap by
exploiting the analogies between the $\Phi \to K^0 \bar{K^0}$ and the singlet EPRB cases. 

In the well--known EPRB configuration one deals with the singlet state 
\begin{equation} \label{2}
|0,0 \rangle = {1 \over \sqrt{2}}\left[|+ \rangle \otimes|- \rangle
-|- \rangle \otimes|+ \rangle \right],
\end{equation}
i.e., an antisymmetric system consisting of two separating components, exactly 
as in eq.(\ref{1}). Also, each one of these two components (say, electrons) is
assumed to have spin--$\frac{1}{2}$ thus belonging to a dimension--two Hilbert space
with basis vectors  $| + \rangle$ and $| - \rangle$, in close analogy to the basis
vectors 
$| K^0\rangle$ and  $|\bar{K^0}\rangle$ in eq.(\ref{1}). In the EPRB
configuration, the experimentalist is supposed to be able to measure 
{\it at will} the spin components along different directions in
both beams, as explicitly required to derive Bell--inequalities in local realistic
contexts. More precisely, we assume that on the left (right) beam one can adjust these
measurement directions either along  $\vec{a}$  or along $\vec{a\prime}$ ($\vec{b}$  or
$\vec{b \prime}$). In the appropriate units, the outcomes of these measurements
are simply $\pm$ signs, i.e., $\sigma_i = \pm$ for $i = \vec{a}, \vec{a \prime}, \vec{b},
\vec{b \prime}$. The probability to obtain specific outcomes (say, $\sigma_a$ and
$\sigma_b$)  when measuring along given directions (say, $\vec{a}$ on the left and 
$\vec{b}$ on the right) will be denoted by 
$P(\vec{a},\sigma_a;\vec{b},\sigma_b)$
and by similar expressions for alternative outcomes and orientations.

In the context of quantum mechanics, all these probabilities can be unambiguosly
computed. For the singlet state one obtains
\begin{eqnarray} \label{3}
& &P(\vec{a}, +;\vec{b}, +) = P(\vec{a}, -;\vec{b}, -) = 
{1 \over 4}(1 - \cos \theta_{ab}) = {1 \over 2} \sin^2 (\theta_{ab}/2) \nonumber \\
& &P(\vec{a}, +;\vec{b}, -) = P(\vec{a}, -;\vec{b}, +) = 
{1 \over 4}(1 + \cos \theta_{ab}) = {1 \over 2} \cos^2 (\theta_{ab}/2) , 
\end{eqnarray}
where $\theta_{ab}$ is the angle between $\vec{a}$ and $\vec{b}$. 
In the context of  local realistic theories, rather than explicitly computing
probabilities, one can establish several Bell--inequalities to be satisfied by these
probabilities in alternative experimental set--ups,  
\begin{equation} \label{wig}
P(\vec{a}, +;\vec{b}, +) \leq P(\vec{a}, +;\vec{c}, \sigma_c) + 
P(\vec{c}, \sigma_c;\vec{b}, +), 
\end{equation}
where $\sigma_c$ can be either + or -- and $\vec{c}$ stands for a given direction 
common to both left ($\vec{c} = \vec{a \prime}$) and right ($\vec{c} = \vec{b \prime}$) 
hand sides. This is the Wigner version of the Bell--inequality and can easily be derived
(for details, see \cite{wigner},\cite{review1},\cite{domenico}) for deterministic,
local hidden--variable theories. It holds for the most interesting case in which one
has space--like  separation between the left and right spin--measurement events. This
is simply achieved working in a symmetric configuration, i.e., placing the detectors
at equal (time-of-flight) distances from the origin. Bell's theorem then 
establishes the incompatibilitiy between these theories and quantum mechanics by simply
proving that the probabilities in eq.(\ref{3}) can violate the Wigner--inequality 
(\ref{wig}).

Before entering into this violation, let us reformulate our simple EPRB analysis in a
slightly different configuration. Assume now that the experimentalist is constrained to 
measure spin projections of the spin--$\frac{1}{2}$ particles along a single and fixed
direction common to both left and right hand beams (say, the vertical or
$z$-direction). All the  discussion of the previous paragraph can be maintained if the
experimentalist is allowed to  introduce magnetic field(s) along the electrons
path(s). Indeed, if the magnetic field $B$  is adjusted to produce a rotation of the
spinor around the propagation axis of angle $\theta_{ab} \equiv \theta_B \equiv
\omega_B\Delta t$ on only one of the two electrons, then the same expressions
(\ref{3}) are the correct quantum mechanical predictions  for the different
probabilities to measure the left and right vertical spin--components. This can be
immediately seen substituting the effects of the rotation,
\begin{eqnarray} \label{rot}
|+ \rangle &\to & \cos (\theta_{ab} /2) |+ \rangle + \sin (\theta_{ab} /2) |- \rangle
\nonumber \\
|- \rangle &\to & \cos (\theta_{ab} /2) |- \rangle - \sin (\theta_{ab} /2) |+ \rangle ,  
\end{eqnarray}
in the second kets (say) of eq.(\ref{2}). One then finds 
\begin{eqnarray} \label{rot2}
|0,0 \rangle \to {1 \over \sqrt{2}} \big[ \cos (\theta_{ab} /2) 
(|+ \rangle \otimes|- \rangle - |- \rangle \otimes|+ \rangle ) -& & \nonumber \\
\sin (\theta_{ab} /2) 
(|+ \rangle \otimes|+ \rangle + |- \rangle \otimes|- \rangle ) \big] &,&
\end{eqnarray}
thus recovering the quantum mechanical probabilities in eq.(\ref{3})
\begin{eqnarray} \label{pro2}
& &P(0, +;B, +) = P(0, -;B, -) = 
{1 \over 4}(1 - \cos \theta_{B}) = {1 \over 2} \sin^2 (\theta_{B}/2) \nonumber \\
& &P(0, +;B , -) = P(0, -;B, +) = 
{1 \over 4}(1 + \cos \theta_{B}) = {1 \over 2} \cos^2 (\theta_{B}/2) , 
\end{eqnarray}
with a new notation indicating explicitly the presence of the magnetic field $B$ on the
right and  its absence on the left. In the context of 
local realistic theories, two Wigner--inequalities (\ref{wig}) can be derived,   
\begin{eqnarray} \label{wig2}
P(0, +;2B, +) &\leq& P(0, +;B, +)  + P(B, +;2B, +) \nonumber \\
P(0, +;2B, +) &\leq& P(0, +;B, -)  + P(B, -;2B, +),
\end{eqnarray}
where the first (second) inequality implies $\frac{1}{2} \sin^2 (\theta_{B}) \leq \sin^2 
(\theta_{B}/2)$  ($\frac{1}{2} \cos^2 (\theta_{B}) 
\newline \leq \cos^2 (\theta_{B}/2)$) and   
is violated for rotation angles $0 < \theta_{B} <\pi /2$  ($\pi /2 < \theta_{B} < \pi$).
Care has to be taken, however, to concentrate the magnetic field in a small region just
before detection in such a way that the spin--measurement event on the left  is
space--like separated from the whole rotation interval $\Delta t$ and spin--measurement
on the right. A tiny violation of the first inequality (\ref{wig2}) persists
even for small values of $\theta_B $. 
 
We now turn to the $\Phi \to K^0 \bar{K^0}$ case, where one is really constrained to the
situation of the preceding paragraph. Indeed, only the two basis states 
$|{K^0}\rangle$ and $|\bar{K^0}\rangle$ can be unambiguously identified on both sides  
by means of their distinct strangeness--conserving strong interactions
on nucleons (see \cite{ghi}; see also \cite{abn} for a discussion
on this issue). One then needs to mimic the preceding effects of an adjustable  
magnetic field. A thin, homogeneous slab of ordinary (nucleonic) matter placed just
before one of the two $K^0 \bar{K^0}$--detectors will do the job.  
The effects of this slab --a neutral kaon regenerator with regeneration parameter
$\rho$--  on the entering, freely propagating $|K_{S/L} \rangle $ states are (see
Appendix)
\begin{eqnarray}\label{pseudorot}
|K_{S} \rangle & \to & |K_{S} \rangle  + r |K_{L} \rangle \nonumber \\
|K_{L} \rangle & \to & |K_{L} \rangle  + r |K_{S} \rangle ,
\end{eqnarray}
where only first order terms in the (small) parameter $r$ 
have been kept. Notice the 
strong similarity between $r \equiv (im_S - im_L +\frac{1}{2}\Gamma_S
-\frac{1}{2}\Gamma_L)\times\rho\Delta t$ and  the previous 
$\theta_B \equiv \omega_B\Delta t = (ge\hbar /2m)\times B\Delta t$  in that both
expressions contain a first factor characterizing the  propagating particles times a
second one allowing for different choices of external
intervention. But notice also that the transformation (\ref{pseudorot}) with a
complex $r$ is not  a true rotation in contrast to (\ref{rot}). Introducing the
regenerator on the right beam, as before, and  finally reverting to the  $K^0
\bar{K^0}$--basis, eq.(\ref{1prime}) becomes 
\begin{eqnarray} \label{1rot}
|\Phi (0) \rangle \to |\Phi (t) \rangle &\simeq& {N(t) \over \sqrt{2}}
\left[(1-r)|K^0\rangle \otimes|\bar{K^0}\rangle -
(1+r)|\bar{K^0}\rangle \otimes|K^0 \rangle \right].
\end{eqnarray}
As in the spin case, the antisymmetry of the initial state has been lost although not 
in the same way, as expected from the differences between
(\ref{rot}) and (\ref{pseudorot}).

In the context of quantum mechanics one can unambiguosly compute the  
detection probabilities by simply projecting eq.(\ref{1rot}) over the appropriate states 
\begin{eqnarray} \label{pqm}
P(0,K^0;r,\bar{K^0}) &=& P(r,\bar{K^0};0,K^0) \simeq N(t)/2 - N(t) {\it {Re}} (r)
\nonumber \\ P(0,\bar{K^0};r,K^0) &=& P(r,K^0;0,\bar{K^0}) \simeq N(t)/2 + N(t) 
{\it{Re}} (r) \nonumber \\ 
P(0,\bar{K^0};r,\bar{K^0}) &=& P(r,\bar{K^0};0,\bar{K^0}) \simeq 0 \nonumber \\
 P(0,{K^0};r,K^0) &=& P(r,{K^0};0,K^0) \simeq 0, 
\end{eqnarray}
where the left equalities are an obvious consequence of rotation invariance and 
the approximated ones in the right are valid at first order in $r$. The notation 
follows closely that in eq.(\ref{pro2}) with the  
$K^0$ or $\bar{K^0}$ indicating the outcome of the measurement and $r$ or $0$  indicating
the presence or absence of the regenerator. Under the same conditions as before, one can
now invoke  local realistic theories to establish Wigner--inequalities such as   
\begin{eqnarray}\label{wigk}
P(0,K^0;0,\bar{K^0}) \leq P(0,{K^0};r,K^0) +  P(r,K^0;0,\bar{K^0}) \nonumber \\
P(0,K^0;0,\bar{K^0}) \leq P(0,{K^0};r,\bar{K^0}) +  P(r,\bar{K^0};0,\bar{K^0}).
\end{eqnarray}
The incompatibility between quantum mechanics and local realism appears when  
introducing the probabilities (\ref{pqm}) in (\ref{wigk}): the first inequality leads
to  
${\it {Re}} (r) \geq 0$, while the second one leads to ${\it {Re}} (r) \leq 0$.
Hence, in any case (i.e., independently of the specific properties of
the regenerator) one of the Wigner--inequalities (\ref{wigk}) is violated by 
quantum mechanics.

This result contrasts with the previously mentioned ones coming from early  
attempts to check local realistic theories in $e^+e^- \to \Phi \to K^0 \bar{K^0}$ 
\cite{datta1}-\cite{ghi}, where interesting Bell--inequalities involving different 
$K^0 \bar{K^0}$ detection {\it times} were proposed. Choosing among these different 
times entails the required active intervention of the experimentalist (as 
particularly emphazised in 
\cite{bigi}), but the inequalities so derived failed in showing their violation by
quantum mechanics due to the specific values of kaon masses and widths. 
More recently, there has been a renewed interest in this subject 
\cite{domenico}-\cite{uchiyama} but, in spite of several claims, 
we believe that the proposed Bell--inequalities do not follow strictly  
from local realism. Indeed, detection of kaonic states other than 
$K^0 ,\bar{K^0}$ is required and their identification via their associated decay 
modes is proposed (see \cite{abn} for details). 
But the simple observation and counting of these decay events offer no option for
an active intervention of the experimentalist, as required to establish 
these inequalities in a local realistic context. 
This clearly contrasts with our
proposal, where freely adjustable regenerators are involved in alternative 
experimental set--ups.  
Finally, we would like also to emphasize that in spite of certain
analogies the
$K^0-\bar{K}^0$ system displays interesting differences as compared to the usually
considered photons or electrons. 
Indeed, the $K^0-\bar{K}^0$ system has some unique and 
peculiar quantum mechanical properties: it is unique as it is 
the only place in nature where CP--violation has been detected so far; 
it is
peculiar since the non--mass eigenstates are unstable and manifest  $K^0 \bar{K^0}$ 
oscillations in space--time. This could add some relevance to our results, 
which, on the other hand, require further analyses aiming to increase the  
tiny violation effects encountered here to a higher, fully 
observable level. 
\vskip 1.5cm

{\bf Appendix}

We define the CP $=\pm 1$ eigenstates $K_{1/2}$ by 
$|K_{1/2}\rangle ={1 \over \sqrt{2}}\left[|K^0 \rangle \pm |\bar{K^0}\rangle \right]$. 
The mass eigenstates $K_{S/L}$ in terms of $K_{1/2}$ and the CP 
violation parameter $\epsilon$ are
\begin{eqnarray} \label{sl}
|K_S\rangle &=&{1 \over \sqrt{1+|\epsilon|^2}}\left[|K_1\rangle
+\epsilon |K_2 \rangle \right] \nonumber \\
|K_L\rangle &=&{1 \over \sqrt{1+|\epsilon|^2}}\left[|K_2\rangle
+\epsilon |K_1 \rangle \right] .
\end{eqnarray}
The time development of these non--oscillating mass eigenstates
is given by $|K_{S/L}(t)\rangle = e^{-i\lambda_{S/L}t}|K_{S/L}\rangle $, with 
$\lambda_{S/L} \equiv m_{S/L}-{i \over 2}\Gamma_{S/L}$, and 
$m_{S/L}$ and $\Gamma_{S/L}$ being the mass and width of $K_S$ and
$K_L$, respectively.  

For kaon regeneration in homogeneous nucleonic media 
we follow \cite{domenico}, \cite{ab} and \cite{kabir}. The eigenstates of the mass matrix
inside matter are
\begin{eqnarray} \label{46}
|K_S'\rangle \simeq |K_S \rangle -\varrho |K_L \rangle
\nonumber \\
|K_L'\rangle \simeq |K_L \rangle +\varrho |K_S \rangle,
\end{eqnarray}
where only first order terms in $\varrho$ have been retained. 
This regeneration parameter is 
$\varrho = \pi \nu (f-\bar{f}) / m_K (\lambda_S-\lambda_L)$,
where $m_K=(m_S + m_L)/2$, $f$($\bar{f}$) is the forward scattering
amplitude for $K^0$($\bar{K}^0$) on nucleons and $\nu$ is the nucleonic density.
This is probably the easiest parameter to adjust in an experimental 
set--up. The time evolution inside matter for the eigenstates $|K_{S/L}'\rangle$ 
follows the standard exponential form,  
$|K_{S/L}'(t)\rangle =e^{-i\lambda_{S/L}'t}|K_{S/L}'\rangle $, where
$\lambda_{S/L}' = \lambda_{S/L}- \pi \nu (f+\bar{f}) /m_K +{\cal O}(\varrho^2)$.  
To compute the net effect of a thin regenerator slab over the 
entering  $|K_{S/L}\rangle$ states one simply expresses these states in the 
$|K_{S/L}'\rangle$ basis using (\ref{46}), introduces their time evolution in 
$\Delta t$ and reverts to the initial $|K_{S/L}\rangle$ basis (see, for
instance, \cite{domenico}  and \cite{kabir}). One finds 
\begin{eqnarray} \label{50}
|K_S\rangle &\to& e^{-i\lambda_S'\Delta t}
\left(|K_S \rangle +i \varrho (\lambda'_S - \lambda'_L )\Delta t |K_L \rangle \right)
\simeq |K_S \rangle + r |K_L \rangle \nonumber \\
|K_L\rangle &\to& e^{-i\lambda_L'\Delta t}
\left(|K_L \rangle +i \varrho (\lambda'_S - \lambda'_L )\Delta t |K_S \rangle \right)
\simeq |K_L \rangle + r |K_S \rangle \, ,
\end{eqnarray}
where $\Delta t$ is short enough to justify 
the systematic use of first order approximations. Eq. (16) defines
the parameter $r$ entering equations (10)--(13); in a first approximation
we have  $r \simeq i (\lambda_S -\lambda_L )\times\rho \Delta t$, as quoted in the
main text.
 
\vskip 0.5cm
{\bf Acknowledgments}. This work has been partly 
supported by the Spanish Ministerio de Educaci\'on y Ciencia, by the 
Funda\c{c}\~ao de Amparo \`a Pesquisa do Estado de S\~ao Paulo (FAPESP) 
and Programa de Apoio a N\'ucleos de Excel\^encia (PRONEX), 
and by the EURODAPHNE EEC-TMR program CT98-0169. Discussions with 
R. Mu\~noz-Tapia are also acknowledged.


\begin{thebibliography}{99}
\bibitem{remark1}
For reasons of simplicity and to maintain space--like separation
between the events of the left and right hand side, we restrict
ourselves to equal--time development. This has also the obvious
advantage of maximal entanglement for all times. Note also 
that for non--equal time evolution the time dependence does not
factorise as in (2).
\end{thebibliography}
\end{document}